# Deep Learning-Based Automated Post-Operative Gross Tumor Volume Segmentation in Glioblastoma Patients

Rajarajeswari Muthusivarajan [1], Adrian Celaya [1,4], Maguy Farhat [2], Wasif Talpur [2], Holly Langshaw [2], Victoria White [2], Andrew Elliott [2], Sara Thrower [2], Dawid Schellingerhout [3], David Fuentes [1,*] and Caroline Chung [2,*]

[1]Department of Imaging Physics, University of Texas MD Anderson Cancer Center, Houston, TX 77030 USA.
[2]Department of Radiation Oncology, University of Texas MD Anderson Cancer Center, Houston, TX 77030 USA.
[3]Department of Neuroradiology, Division of Diagnostic Imaging, University of Texas MD Anderson Cancer Center, Houston, TX 77030 USA.
[4]Department of Computational Applied Mathematics and Operations Research, Rice University, Houston, TX 77005 USA.

***Correspondence:** cchung3@mdanderson.org and dtfuentes@mdanderson.org

**Abstract:** Precise automated delineation of post-operative gross tumor volume in glioblastoma cases is challenging and time-consuming owing to the presence of edema and the deformed brain tissue resulting from the surgical tumor resection. The pre-operative glioma segmentation models that have been developed to date struggle to define the post-operative cavity and any residual tumor; however, the post-operative tumor and cavity are what are tracked to assess treatment response. To develop a model for automated delineation of post-operative gross tumor volumes in glioblastoma, we proposed a novel 3D double pocket U-Net architecture that has two parallel pocket U-Nets. Both U-Nets were trained simultaneously with two different subsets of MRI sequences and the output from the models was combined to do the final prediction. We strategically combined the MRI input sequences (T1, T2, T1C, FL) for model training to achieve improved segmentation accuracy. The dataset comprised 82 post-operative studies collected from 23 glioblastoma patients who underwent maximal safe tumor resection. All had gross tumor volume (GTV) segmentations performed by human experts, and these were used as a reference standard. The results of 3D double pocket U-Net were compared with baseline 3D pocket U-Net models and the ensemble of 3D pocket U-Net models. All the models were evaluated with fivefold cross-validation in terms of the Dice similarity coefficient and Hausdorff distance. Our proposed double U-Net model trained with input sequences [T1, T1C, FL + T2, T1C] achieved a better mean Dice score of 0.8585 and Hausdorff distance of 4.1942 compared to all the baseline models and ensemble models trained. The presence of infiltrating tumors and vasogenic edema in the post-operative MRI scans tends to reduce segmentation accuracy when considering the MRI sequences T1, T2, T1C, and FL together for model training. The double U-Net approach of combining subsets of the MRI sequences as distinct inputs for model training improves segmentation accuracy by 7% when compared with the conventional method of model training with all four sequences.

**Keywords:** Glioblastoma; Deep learning; Gross tumor volume segmentation; Post-treatment MRI; convolutional neural networks

## 1. Introduction

Glioblastoma is an aggressive malignant primary brain tumor that is typically treated with maximal safe surgical resection followed by radiation treatment and chemotherapy [1]. Because glioblastomas have an infiltrating growth pattern, complete resection of all the tumor cells is not possible. A crucial step for precise radiation treatment is the delineation of the treatment target. The gross tumor volume (GTV) is defined as the surgical cavity along with the residual enhancing tumor on contrast-enhanced T1-weighted (T1C) post-operative MRI scan [2]. Manual delineation of the GTV is a labor-intensive, time-consuming process [3] and is prone to intraobserver and interobserver variability [4-6]. The development of a reliable and robust automated segmentation tool could serve to improve the consistency and efficiency of radiation treatment planning while also serving to help evaluate disease response or progression in follow-up and provide consistent segmentation tools for the development of predictive biomarkers.

With recent advancements in convolutional neural networks, several automated segmentation models have been developed for gliomas. These models were primarily trained with pre-operative MRI scans of patients with glioblastoma for tumor segmentation, including enhancing tumor, peritumoral edema, necrotic, and non-contrast-enhancing tumors [7-9]. The deep learning architecture U-Net [10] and its variants DenseNet [11], ResNet [12], DeepMedic [13] and nnU-Net [14] have shown promising tumor segmentation accuracy with a Dice score of 0.9 [15, 16] and outperform the traditional algorithms. However, models trained with pre-operative MRI scans may not be suitable for predicting segmentations with post-operative scans [17]. The post-operative MRI scans often include swelling, bleeding, and inflammations in the brain tissue due to the surgical resection which makes the segmentation more challenging.

One alternative solution to this is to fine-tune the pre-trained model with a transfer learning approach. Ghaffari et al [17] demonstrated the automated segmentation task on post-operative MRIs by fine-tuning the model trained on pre-operative MRIs from BraTS 2020 dataset. They utilized the transfer learning approach to compensate for the small size of the post-operative data. Bianconi et al [18] followed a similar approach to a segment glioblastoma on both pre-operative and post-operative MRIs. They used a model pre-trained on BraTS2021 data and attained a mean Dice score of 0.6352 for the resection cavity and 0.7231 for the resection cavity, GTV, and whole tumor together. However, the segmentation accuracy for the post-operative scans still falls behind compared to the pre-operative scans.

Several semi-automated approaches were proposed based on voxel or shape-based methods such as fuzzy algorithms [19], level-set approaches [20, 21] hybrid generative discriminative framework [22]. Along with MRI, CT images were also used for the automated delineation of GTVs [23, 24]. There are only a handful of reports published on fully automated segmentation of the resection cavity or GTV on post-operative MRI scans of glioblastoma [25-28]. Helland et al [26] demonstrated automated segmentation of residual enhancing glioblastoma on a multicenter dataset which comprised early post-operative MRI scans T1, T1C, FL (No T2), and pre-operative T1C (acquired within 72 hours after surgery). The model trained with post-operative T1C and T1 achieved a 0.6161 Dice score with nnU-Net and 0.5314 with AGU-Net. Including post-operative FL or pre-operative T1C in the training decreased the model performance.

To the best of our knowledge, Ermis et al [27] were the first to study fully automated resection cavity segmentation only using post-operative MRI scans. They used a DenseNet architecture to process MRIs in three separate sets of 2D plane-wise orientations (axial, coronal, and sagittal), resulting in three separate predictions. The average of the three predictions gave the final 3D volume of the resection cavity, which they compared with the resection cavity ground truths defined on the T1 and T2 images by three radiation oncologists. The reported median Dice scores of the automated segmentation compared to the three radiation oncologists were 0.83, 0.81, and 0.81.

Very recently, Ramesh et al. [28] developed a lighter-weight model using only two post-operative image sequences FL and T1C with and without skull stripping, for segmenting the GTV1, which was defined as the resection cavity including the hyperintense lesion cavity and blood products, and GTV2, which was defined as the resection cavity and residual enhancing tumor. Their best-performing 3D U-

Net model resulted in mean Dice scores of 0.72 and 0.73 for the GTV1 and GTV2 segmentations, respectively. As far as we know, that was the only attempt to segment GTVs in patients with glioblastoma. However, this study excluded T1 and T2 images in the model training and relied on FL and T1C for both GTV1 and GTV2 segmentations.

A current gap in knowledge is how effectively the different imaging modalities can be utilized for the precise automated segmentation of GTV in post-operative MRI scans of glioblastoma. Previous models developed for post-operative glioblastoma segments residual disease or resection cavity, or studies do not include all four MRI modal images. The goal of our study was to develop a fully automated segmentation model for post-operative GTVs by exploring the role of different MR imaging modalities in segmentation accuracy. To address this, we propose an innovative approach with a double U-Net architecture which will have two independent U-Nets and can take two different inputs for model training. The outputs of the individual U-Nets were combined and further processed through a convolution layer before prediction. In general, researchers have used pre-trained models to build a stacked neural network architecture [29]. In our approach, we trained two independent U-Nets simultaneously with two different inputs to explore which input combination of MR sequences best informed the definition of the post-operative GTV. For that, first, we developed eight baseline models by training the neural network (3D Pocket U-Net [30]) with subsets of MRI volumes rather than training with all four modalities together. Based on the performance of the baseline models, we strategically selected the input pairs for our double U-Net architecture. The model can thus capture the modality-specific information from the given input pairs and retain the features extracted to predict more effectively. This technique can help improve generalization and robustness by leveraging diverse models. For comparison, we also performed ensemble modeling [31-33] with the pre-trained baseline models. We obtained the final GTV prediction of the ensemble model (EM) via model averaging.

## 2. Materials and Methods
### 2.1. Dataset
The dataset contains 82 longitudinal multimodal MRI post-operative scans collected from 23 patients (Table 1). All are identified with glioblastoma and underwent maximal safe tumor resection followed by standard radiation therapy at our Institution. From the same patient, scans were collected at multiple time points, and each was considered as an independent sample. The post-operative MRI scans were performed between 1 to 354 days after the surgery date and the median time point for the post-operative MRI is 155 days. The data of each patient consists of native (T1), T2-weighted (T2), post-contrast T1-weighted (T1C), and T2-Fluid attenuated inversion recovery (FL) MRI brain scans.

**Table 1.** Demographic information of patients considered for this study.

| Variable | Value (Median) |
|---|---|
| Total number of patients | 23 |
| Total number of samples | 82 |
| Imaging taken after post-surgery | 1-354 days (155) |
| Number of scans per patient | 1-8 (4) |

### 2.2. Delineation of ground truth
For each case, annotations of GTV (resection cavity along with residual enhancing tumor as one volume) were manually contoured on T1 post-contrast images using semi-automated tools on Raystation treatment planning software (Figure 1). Each contour was reviewed by a radiation oncologist having at least 10 years of experience.

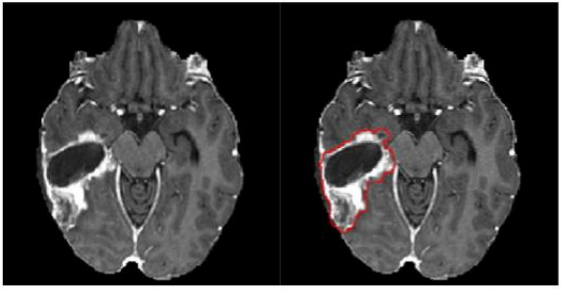

**Figure 1.** An example of a T1C image and manually annotated GTV overlayed on T1C. GTV includes a resection cavity along with the enhancing tumor (shown as a red contour).

## 2.3. Preprocessing

Image registration was performed using the open-source program Advanced Normalization Tools (ANTs) package [34]. All volumes were co-registered to the same anatomical template as the T2 reference image. Brain extraction from non-brain parts (skull stripping) was done with the HD Brain Extraction tool (HD-BET) [35]. Images have been standardized with SimpleITK [36] for isotropic voxel spacing of 0.86 mm$^3$ and interpolated to the same to a resolution of 256 × 256 × 208. Images were Z-score normalized to a mean value of 0 and a standard deviation value of 1.

## 2.4. Network architecture and training

### 2.4.1. Baseline model

PocketNet [30] version of a 3D U-Net was implemented for model training. Like U-Net, Pocket U-Net has a combination of convolution and downsampling operations that extract features along a contracting path and upsamples feature maps with transposed convolution layers in the expanding path. However, the key idea behind PocketNet is that the number of feature maps from each convolution layer remains constant, regardless of the spatial resolution which was shown to reduce the number of learnable parameters without compromising accuracy. In each epoch, 16 randomly positioned $64^3$ patches were extracted from each image and ground truth mask and fed into the network. The learning rate was set to 0.02, the batch size was set to 16, and the number of epochs was set to 200. A stochastic gradient descent with momentum optimization was used. We adopted a fivefold cross-validation scheme where the data were partitioned as follows: 70% for training, 10% for validation, and 20% for testing. The same seed (18) was used for each fivefold model split for a fair comparison of all the models. The hyperparameters kernel size, patch size, batch size, and channels were optimized by comparing the mean Dice scores. All four MRI sequences were used for the parameter sweeping. All our experiments were implemented using the MONAI framework with Pytorch on a workstation with an NVIDIA GeForce RTX 8000 graphics card with 48 GB of RAM. In total, eight baseline models were trained with subsets of MRI volumes, ranging from all four image sequences to fewer, and single sequence inputs: BM[T1, T2, T1C, FL], BM[T1, T2, T1C], BM[T2, T1C, FL], BM[T1, T1C, FL], BM[T1, T1C], BM[T2, T1C], BM[T1C, FL] and BM[T1C]. Because the ground truth GTV mask was defined based on the T1C image, we ensured that all the trained baseline models included the T1C image in one input channel.

### 2.4.2. Double U-Net Model

We propose an innovative approach with a double U-Net architecture with two independent Pocket U-Nets which can take two pairs of inputs for model training. Figure 2 shows the overview of the Double U-Net architecture. Both Pocket U-Nets are trained simultaneously with specific input data, independent of each other. The output from the last decoder of both Pocket U-Nets was channel-wise concatenated to preserve the features learned from each input set. Then it passed through a convolutional block consisting of two 3 x 3 convolutions in which each convolution was followed by batch normalization and rectified linear unit function (ReLU). The output from the convolutional block passed through a 1 x 1 convolutional layer with a softmax activation function providing the final pixel-wise classification.

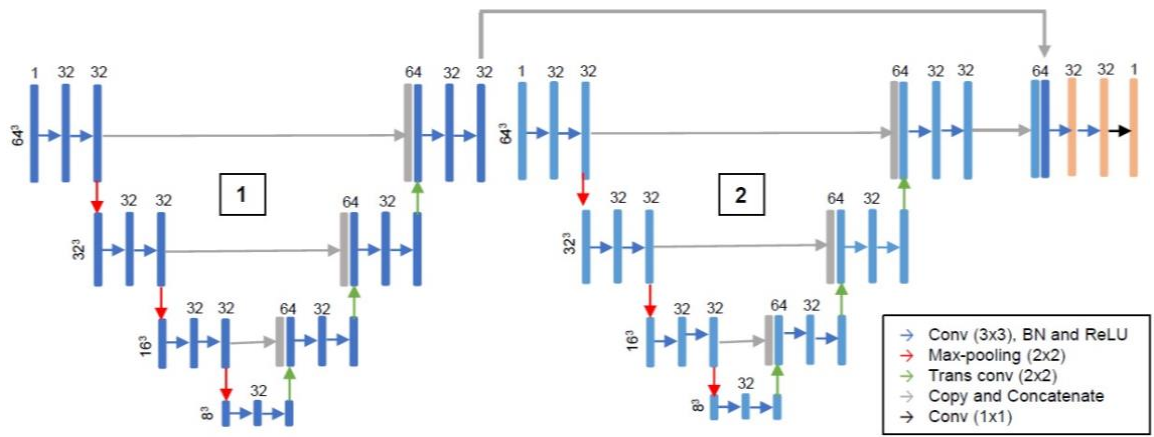

**Figure 2.** The double pocket U-Net architecture in which the output of the last decoders of the individual Pocket U-Nets was concatenated and passed through the double convolutional layer before the pointwise convolutional operation.

### 2.4.3. Ensemble Model

In addition, the model averaging ensemble model (EM) method was performed using the pre-trained baseline models [31, 33]. In this approach, the EM combines the predictions of any two baseline models (both models are given the same weightage) by concatenating the softmax outputs of the models (Figure 3). Concentrating the predictions from multiple models reduces the total error from an individual baseline model, which is expected to increase the overall model performance by introducing diversity.

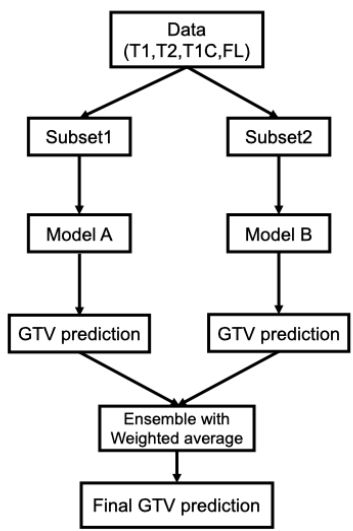

**Figure 3.** Block diagram of weighted average ensemble model for GTV segmentation.

### 2.5. Model Evaluation

The fivefold cross-validation performance of a model was evaluated in terms of Dice similarity coefficient, Hausdorff distance, false positive error, and false negative error as implemented with SimpleITK [36]. Model performance was presented on the held-out test set of each fold. The Dice similarity coefficient [37] between the segmented mask (S) and its corresponding ground truth mask

(T) was calculated using the formula $D = \frac{2 \times |S \cap T|}{|S|+|T|}$ Where S is the set of voxels in the predicted mask, and T is the set of voxels in the ground truth mask. Hausdorff distance is calculated by computing the distance between the set of non-zero pixels of two images (predicted mask and ground truth mask). In this work, this was adapted to yield Hausdorff 95, a metric to measure the 95% quantile of the surface distance. The rate of positive predictions (over-segmentation) that were false positives is defined as a false positive error (FPE). Note that here the false positive error does not include the number of negative voxels and subsequently has no bias to increased field of view or smaller segmentations, a problem that happens with specificity and other implementations of FPE. The rate at which positive voxels were classified as negative is defined as false negative error (FNE) (under segmentation). Wilcoxon signed-rank testing was conducted to determine if a statistically significant pairwise difference existed between the model performances.

$$FNE = \frac{|T \setminus S|}{|T|}$$
$$FPE = \frac{|S \setminus T|}{|S|}$$

## 3. Results

As a first step, we trained baseline models (BM) with different input subsets and compared their performance with a BM model trained with all four sequences (i.e., BM [T1, T2, T1C, FL]). The mean Dice similarity coefficient and Hausdorff distances of the baseline models are shown in Table 2A. Baseline models trained with inputs [T1, T2, T1C], [T2, T1C], and [T1C] had similar mean Dice similarity coefficients of about 0.82 but had slight variations in Hausdorff distance. BM [T1C] had the best Dice similarity coefficient and Hausdorff distance of 0.8181 and 5.3797 respectively compared to the model trained with all four sequences BM [T1, T2, T1C, FL] (Dice similarity coefficient: 0.7885 and Hausdorff distance: 6.2048). To understand how the input sequence used for the model training affects segmentation accuracy, we compared the predictions from the baseline model BM[T1, T2, T1C, FL] with all other baseline models. There are seven studies identified with less than 0.6 Dice similarity coefficient from the model BM[T1, T2, T1C, FL] predictions. Figure 4 shows the comparison of the Dice similarity coefficient and Hausdorff distance of the selected studies between the all-baseline models. It demonstrates the variation in the model performance concerning the input sequence chosen for model training. Some cases had more than about 50% improvement in segmentation accuracy.

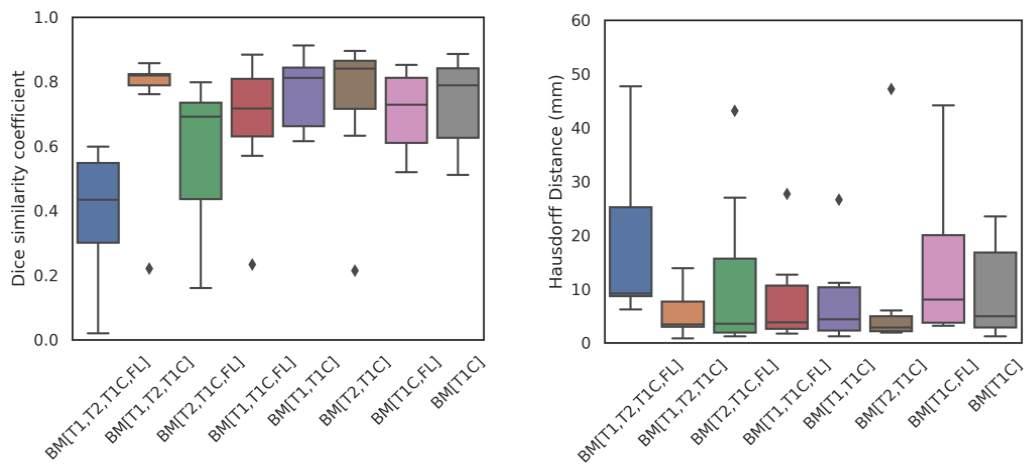

**Figure 4.** Baseline model performance compared over a subset obtained from a model BM [T1, T2, T1C, FL] with less than 0.6 Dice score. The Dice similarity coefficient and Hausdorff distances show the variation in the model accuracy depending on the input sequence.

**Table 2.** The accuracy of the segmentation models (A. Baseline models, B. Double U-Net models C. Ensemble models) was evaluated using the Dice similarity coefficient and Hausdorff 95 distance (mm). [†]To ensure the statistical significance of the model performance, the Wilcoxon signed rank test was conducted with a significance level of 0.05 between the baseline model BM [T1, T2, T1C, FL] and all other models.

A) A comparison of the mean Dice similarity coefficient and Hausdorff 95 distance (mm) baseline models.

| Baseline model input | Dice similarity coefficient | | | Hausdorff 95 (mm) | | |
|---|---|---|---|---|---|---|
| | Mean (Std) | Median | p-value | Mean (Std) | Median | p-value |
| BM [T1,T2,T1C,FL][†] | 0.7885 (0.1536) | 0.8356 | - | 6.2048 (11.7887) | 1.923 | - |
| BM [T1,T2,T1C] | 0.8193 (0.1415) | 0.8561 | *** | 6.3464 (13.3798) | 1.72 | * |
| BM [T1,T1C,FL] | 0.7993 (0.1443) | 0.8484 | NS | 5.3882 (11.1672) | 1.923 | * |
| BM [T2,T1C,FL] | 0.8000 (0.1275) | 0.8386 | NS | 6.0522 (10.7122) | 2.58 | NS |
| BM [T1,T1C] | 0.8088 (0.1266) | 0.8463 | NS | 5.701 (11.12) | 1.923 | NS |
| BM [T2,T1C] | 0.8219 (0.1028) | 0.8445 | NS | 5.6912 (10.2762) | 1.923 | NS |
| BM [T1C,FL] | 0.7982 (0.1305) | 0.8375 | NS | 6.5355 (12.4076) | 2.1066 | NS |
| BM [T1C] | 0.8182 (0.1067) | 0.845 | * | 5.3798 (9.4062) | 2.1066 | NS |

B) A comparison of the mean Dice similarity coefficient and Hausdorff 95 distance (mm) double U-Net models.

| Double U-Net model input | Dice similarity coefficient | | | Hausdorff 95 (mm) | | |
|---|---|---|---|---|---|---|
| | Mean (Std) | Median | p-value | Mean (Std) | Median | p-value |
| DM[T1,T2,T1C+T1C,FL] | 0.8368 (0.0847) | 0.8646 | *** | 4.5145 (8.9798) | 1.72 | *** |
| DM[T2,T1C,FL+T1,T1C] | 0.8073 (0.1509) | 0.8531 | * | 6.5507 (14.7855) | 1.923 | NS |
| DM[T1,T1C,FL+T2,T1C] | 0.8585 (0.0733) | 0.875 | *** | 4.1943 (9.2631) | 1.6048 | *** |
| DM[T1,T1C+T2,T1C] | 0.8127 (0.1065) | 0.8476 | NS | 6.196 (11.1499) | 1.923 | NS |
| DM[T2,T1C+T1C,FL] | 0.8258 (0.1027) | 0.8509 | * | 5.0843 (9.5312) | 1.923 | NS |
| DM[T1C,FL+T1,T1C] | 0.7866 (0.1536) | 0.8369 | NS | 7.4456 (14.9467) | 2.58 | NS |
| DM[T1,T1C+T1C] | 0.8306 (0.0948) | 0.8612 | ** | 4.6541 (8.1908) | 1.923 | * |
| DM[T2,T1C+T1C] | 0.8214 (0.1423) | 0.8659 | ** | 5.5941 (12.8487) | 1.923 | NS |
| DM[T1C,FL+T1C] | 0.8336 (0.1019) | 0.8623 | ** | 4.4401 (7.7694) | 1.923 | NS |

C) A comparison of the mean Dice similarity coefficient and Hausdorff 95 distance (mm) of ensemble models.

| Ensemble model input | Dice similarity coefficient | | | Hausdorff 95 (mm) | | |
|---|---|---|---|---|---|---|
| | Mean (Std) | Median | p-value | Mean (Std) | Median | p-value |
| EM[T1,T2,T1C+T1C,FL] | 0.828 (0.1319) | 0.8682 | *** | 6.3014 (12.9507) | 1.72 | ** |
| EM[T2,T1C,FL+T1,T1C] | 0.8429 (0.069) | 0.861 | *** | 4.4864 (8.65) | 1.72 | *** |
| EM[T1,T1C,FL+T2,T1C] | 0.8412 (0.0862) | 0.8691 | *** | 4.6688 (9.5092) | 1.72 | *** |
| EM[T1,T1C+T2,T1C] | 0.8471 (0.0718) | 0.8675 | *** | 4.3273 (8.7531) | 1.72 | *** |
| EM[T2,T1C+T1C,FL] | 0.8378 (0.0827) | 0.859 | *** | 4.9974 (9.6825) | 1.72 | *** |
| EM[T1C,FL+T1,T1C] | 0.833 (0.0909) | 0.866 | *** | 5.273 (10.7446) | 1.72 | * |
| EM[T1,T1C+T1C] | 0.8281 (0.1166) | 0.866 | *** | 5.1051 (10.8389) | 1.72 | ** |
| EM[T2,T1C+T1C] | 0.8457 (0.0702) | 0.8587 | *** | 4.7648 (9.5701) | 1.72 | *** |
| EM[T1C,FL+T1C] | 0.8397 (0.0754) | 0.8566 | *** | 4.6837 (9.2522) | 1.72 | *** |

NS– not significant, p-value > 0.05; *, p-value < 0.05; **, p-value < 0.01; ***, p-value < 0.001.

Each double U-Net model (DM) has two Pocket U-Net which were trained simultaneously with different input sets. In total, nine double U-Net models are trained. Based on the model input, we grouped the trained models into three categories, Group 1: considered all 4 MRI sequences DM [T1, T2, T1C + T1C, FL], DM [T2, T1C, FL + T1, T1C], DM [T1, T1C, FL + T2, T1C], Group 2: considered 3 MRI sequences DM [T1, T1C + T2, T1C], DM [T2, T1C + T1C, FL], DM [T1C, FL + T1, T1C], Group 3: considered 2 MRI sequences DM [T1, T1C + T1C], DM [T2, T1C + T1C], DM [T1C, FL + T1C]. The Dice similarity coefficient and Hausdorff distances calculated were shown in Table 2B. Overall, we saw an increase in the mean Dice similarity coefficient of about ~ 3 to 7 % compared to the baseline models. The DM [T1, T1C, FL + T2, T1C] outperforms all other double U-Net and baseline models with a Dice score of 0.8585 and Hausdorff distance of 4.1942. We also performed ensemble modeling (EM) with pre-trained baseline models for the same input combinations as Double U-Net models (Table 2C). The models EM [T2, T1C, FL + T1, T1C], EM [T1, T1C, FL + T2, T1C], EM [T1, T1C + T2, T1C], and EM [T2, T1C + T1C] attained the similar Dice score of about 0.84, with slight differences in their Hausdorff distances. Also, they exhibit smaller standard deviations exhibiting lesser variation in Dice score as well as Hausdorff distance.

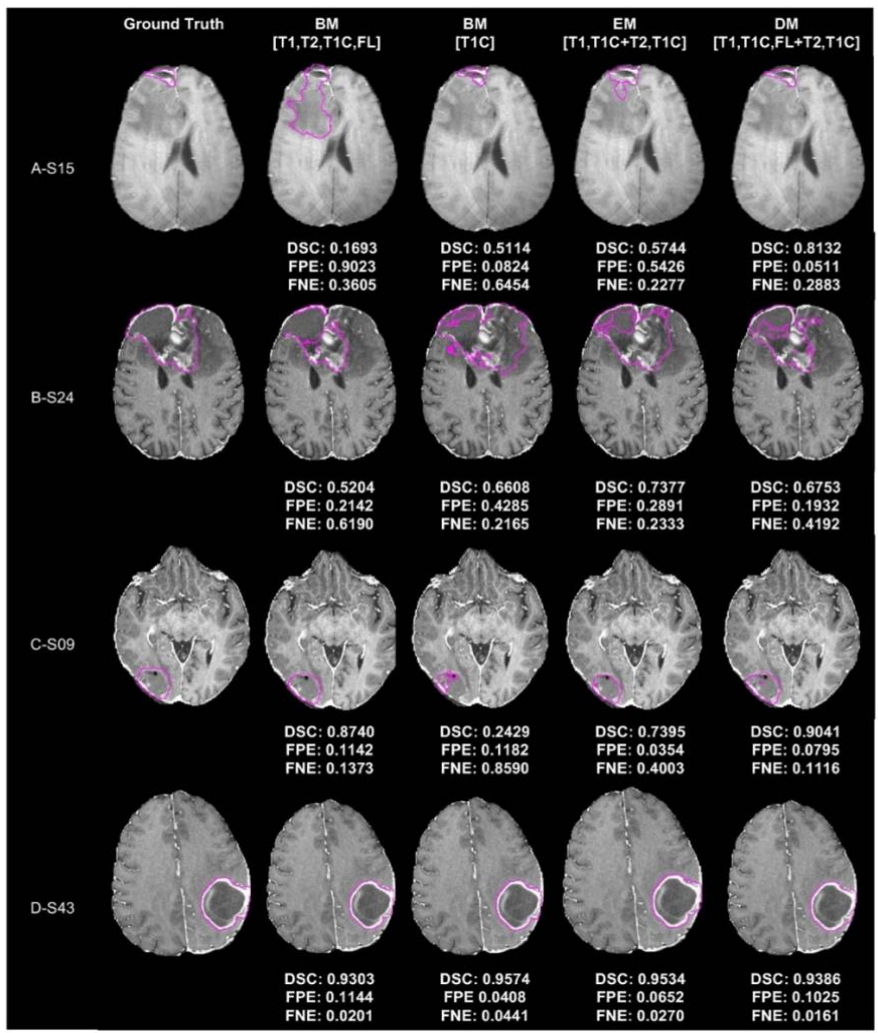

**Figure 5.** Ground truth masks and best-performing model predictions overlayed on the T1C image (all masks shown as pink contour lines). Left to right: Representative studies demonstrating the performance of Baseline models BM [T1, T2, T1C, FL] and BM [T1C], ensemble model, EM [T1, T1C + T2, T1C] and double U-Net model, DM [T1, T1C, FL + T2, T1C] with their Dice similarity coefficient (DSC), False positive error (FPE) and False negative error (FNE).

Figure 5 shows a visual comparison of GTV predictions overlayed on T1C images from the representative studies S15, S24, S09, and S43. Left to the right shows the ground truth, predictions of baseline models, BM [T1, T2, T1C, FL] and BM [T1C], EM [T1, T1C + T2, T1C], and DM [T1, T1C, FL + T2, T1C] and their respective Dice similarity coefficient (DSC), false positive error (FPE) and false negative error (FNE). In, Figure 5a, S15 BM [T1, T2, T1C, FL] overestimated the resection cavity with edema owing to the hyperintense signal from T2 and FL. Removing the T2 and FL sequences from model training helped the model BM [T1C] to eliminate the hyperintense signal which reduces the FPE. At the same time, this makes the model underestimate the resection cavity. However, when considering T2 and FL as separate model inputs as in DM [T1, T1C, FL + T2, T1C], the model attained the best segmentation accuracy (DSC: 0.8132).

Figure 5b, S24 shows average performance across all models. Figure 5c, S09 exhibited a better Dice score when the model was trained with all four modal input sequences. Eliminating one or more sequences from the model training (BM [T1C], EM [T1, T1C + T2, T1C]) makes the model underestimate the resection cavity. On the other hand, Figure 5d S43 was an example of a resection cavity present with a well-defined boundary of residual enhancing tumor. All the models exhibited consistent performance with a Dice similarity coefficient higher than 0.9.

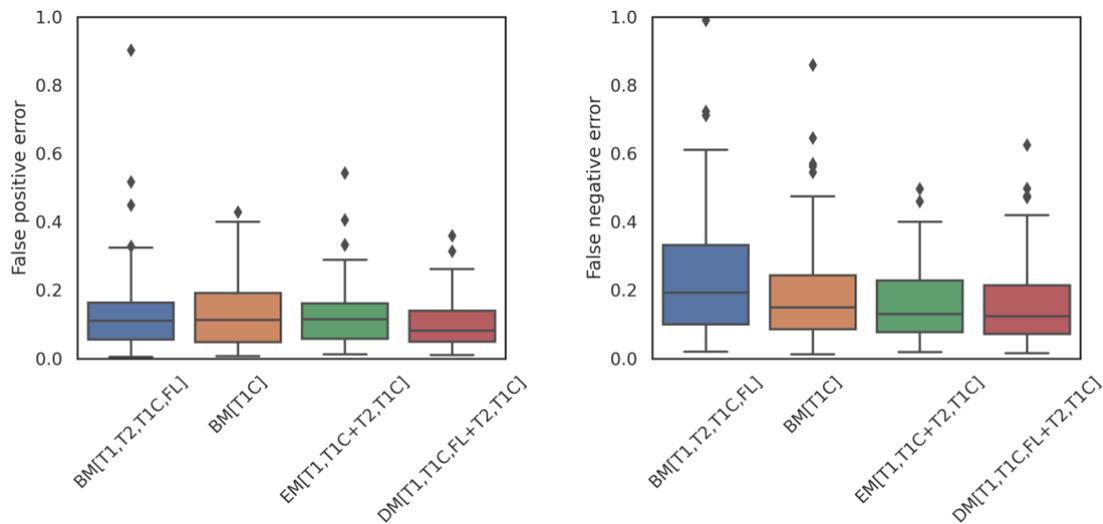

**Figure 6.** Comparison of the false positive error and false negative error of the four best performing models BM [T1, T2, T1C, FL] and BM [T1C], EM [T1, T1C + T2, T1C] and DM [T1, T1C, FL + T2, T1C]. A higher false positive error and a false negative error represent worse performance.

Figure 6 summarizes the distribution of false positive error and false negative error for the four best-performing models BM [T1, T2, T1C, FL], BM [T1C], EM [T1, T1C + T2, T1C], and DM [T1, T1C, FL + T2, T1C]. Overall, the models had higher false negative errors compared to false positive errors. False positive errors arose either from the studies present with edema or when the models tended to over-segment the ventricle as a cavity. We noted that both false positive errors and false negative errors were reduced with the double U-Net model approach which improves the overall segmentation accuracy.

## 4. Discussions
In this study, we evaluated our unique double U-Net architecture for automated segmentation of the post-operative glioblastoma GTV while also investigating the impact of using different subsets of MR sequences as model input data on model performance. Our approach successfully identified the GTV in the presence of edema and other deformations in brain tissue on MRI scans of glioblastoma patients after tumor resection.

Unresected tumor segmentation models with pre-operative MRI scans benefited when the model was trained with all four MR sequences (T1, T2, T1C, FL) [15, 16]. The characteristic information from the different image sequences improved the efficiency of model learning and led to better segmentation accuracy. However, our results from the baseline models demonstrated that the model performance varied based on which input sequences were used for model training. This meant that the best segmentation accuracy could be achieved when combining accurate, diverse individual base models. To this end, we proposed the double U-Net architecture with which we trained two independent U-Nets simultaneously with different input sequences and then combined the extracted features to predict the GTV. Our model performance is validated through a fivefold cross-validation with a held-out test set. The best performing Double U-Net model [T1, T1C, FL + T2, T1C] achieved a Dice score of 0.8585 and a Hausdorff 95 distance of 4.1942.

On the post-operative scans, T1C and T2 sequences showed the soft tissue/lesion with high contrast. However, some gliomas can have non-enhancing, infiltrative components, which are less visible in T1C. FL scans can delineate the infiltrative components in low-grade gliomas (LGG). However, both infiltrating tumors and vasogenic edema in glioblastoma were hyperintense on FL [38]. For automated segmentation of resection cavity volumes, Ermis et al used the 2D approach with all four volumes [27]. In this study, the predicted resection cavity volumes were smaller than the ground truth volumes owing to the signal inhomogeneity from T2 and FL sequences and the presence of other intensity patterns, such as edema, subarachnoid space, or ventricles. This model misinterprets the hyperintense signal arising from the edema in T2 and FL sequences. Ramesh et al observed similar behavior with their lightweight model trained only with T1C and Fl sequences [28] which misinterprets the ventricle for cavity when hyperintense signal arising from FL edema in the same region. This lightweight model underestimates the GTV1 volumes for about 25% of the test cases and achieves an overall mean Dice score of 0.72.

Our baseline models give the intuition for undertaking the hyperintense signals arising from the T2 and FL sequences. BM [T1, T2, T1C, FL] predicted the GTV for S15 along with edema but when the model was trained only with T1C model segmented the GTV with better segmentation accuracy (i.e., small false positive error) by eliminating edema (Figure 5a). However, not all cases benefit from removing T2 and FL sequences from model training. For example, in S09 (Figure 5b), BM [T1C] had an increased false negative error (under segmentation of resection cavity) after removing T2 and FL sequences from model training. Our double U-Net approach improved the model efficiency and overall segmentation performance by eliminating the hyperintense signals from T2 and FL sequences owing to pathological alterations in the surrounding tissues. This robust accurate GTV segmentation aids clinicians in assessing disease progression and in planning follow-up treatment.

## 5. Conclusions

In this study, we showed that high-quality automated GTV delineation in post-operative glioblastoma with a mean Dice similarity coefficient of 0.8585 and Hausdorff 4.1942 can be achieved with our proposed double U-Net model. With the double-U-Net architecture, the modality-specific information from the given inputs was retained and it eliminated the error introduced by signal inhomogeneities that arose from training with multimodal model inputs. The false positive and false negative errors introduced by deformations in brain tissue were reduced with our approach.

review and editing, M.R., A.C., D.S., D.F., C.C.; All authors have read and agreed to the published version of the manuscript.

**Funding:** This work was supported through an Institutional Research Grant (IRG) Program and the MD Anderson Strategic Research Initiative Development (STRIDE) Program – Tumor Measurement Initiative. This work was supported in part by the institutional research grant at MD Anderson and NSF support under Award NSF-2111147 is gratefully acknowledged. We thank the Cancer prevention and Research Institute of Texas (CPRIT) and the Dr. Marnie Rose Foundation for their support. The Department of Defense supports Adrian Celaya through the National Defense Science & Engineering Graduate Fellowship Program.

**References**

[1] F. Hanif, K. Muzaffar, K. Perveen, S. M. Malhi, and U. Simjee Sh, "Glioblastoma Multiforme: A Review of its Epidemiology and Pathogenesis through Clinical Presentation and Treatment," (in eng), Asian Pac J Cancer Prev, vol. 18, no. 1, pp. 3-9, Jan 1 2017, doi: 10.22034/apjcp.2017.18.1.3.

[2] M. Niyazi et al., "ESTRO-EANO guideline on target delineation and radiotherapy details for glioblastoma," Radiotherapy and Oncology, vol. 184, 2023, doi: 10.1016/j.radonc.2023.109663.

[3] P. Y. Bondiau et al., "Atlas-based automatic segmentation of MR images: Validation study on the brainstem in radiotherapy context," (in English), Int J Radiat Oncol, vol. 61, no. 1, pp. 289-298, Jan 1 2005, doi: 10.1016/j.ijrobp.2004.08.055.

[4] M. Visser et al., "Inter-rater agreement in glioma segmentations on longitudinal MRI," Neuroimage Clin, vol. 22, p. 101727, 2019, doi: 10.1016/j.nicl.2019.101727.

[5] G. P. Mazzara, R. P. Velthuizen, J. L. Pearlman, H. M. Greenberg, and H. Wagner, "Brain tumor target volume determination for radiation treatment planning through automated MRI segmentation," (in English), Int J Radiat Oncol, vol. 59, no. 1, pp. 300-312, May 1 2004, doi: 10.1016/j.ijrobp.2004.01.026.

[6] M. A. Deeley et al., "Comparison of manual and automatic segmentation methods for brain structures in the presence of space-occupying lesions: a multi-expert study," Phys Med Biol, vol. 56, no. 14, pp. 4557-77, Jul 21 2011, doi: 10.1088/0031-9155/56/14/021.

[7] B. H. Menze et al., "The Multimodal Brain Tumor Image Segmentation Benchmark (BRATS)," IEEE Trans Med Imaging, vol. 34, no. 10, pp. 1993-2024, Oct 2015, doi: 10.1109/TMI.2014.2377694.

[8] D. Bouget et al., "Glioblastoma Surgery Imaging-Reporting and Data System: Validation and Performance of the Automated Segmentation Task," Cancers (Basel), vol. 13, no. 18, Sep 17 2021, doi: 10.3390/cancers13184674.

[9] H. G. Pemberton et al., "Multi-class glioma segmentation on real-world data with missing MRI sequences: comparison of three deep learning algorithms," Sci Rep, vol. 13, no. 1, p. 18911, Nov 2 2023, doi: 10.1038/s41598-023-44794-0.

[10] O. Ronneberger, P. Fischer, and T. Brox, "U-Net: Convolutional Networks for Biomedical Image Segmentation," in Medical Image Computing and Computer-Assisted Intervention – MICCAI 2015, Cham, N. Navab, J. Hornegger, W. M. Wells, and A. F. Frangi, Eds., 2015// 2015: Springer International Publishing, pp. 234-241.

[11] G. Huang, Z. Liu, L. V. D. Maaten, and K. Q. Weinberger, "Densely Connected Convolutional Networks," in 2017 IEEE Conference on Computer Vision and Pattern Recognition (CVPR), 21-26 July 2017 2017, pp. 2261-2269, doi: 10.1109/CVPR.2017.243.

[12] K. He, X. Zhang, S. Ren, and J. Sun, "Identity Mappings in Deep Residual Networks," in European Conference on Computer Vision, 2016.

[13] K. Kamnitsas et al., "Efficient multi-scale 3D CNN with fully connected CRF for accurate brain lesion segmentation," Med Image Anal, vol. 36, pp. 61-78, Feb 2017, doi: 10.1016/j.media.2016.10.004.

[14] F. Isensee, P. F. Jaeger, S. A. A. Kohl, J. Petersen, and K. H. Maier-Hein, "nnU-Net: a self-configuring method for deep learning-based biomedical image segmentation," Nature Methods, vol. 18, no. 2, pp. 203-211, 2021/02/01 2021, doi: 10.1038/s41592-020-01008-z.

[15] R. Yousef et al., "U-Net-Based Models towards Optimal MR Brain Image Segmentation," (in eng), Diagnostics (Basel), vol. 13, no. 9, May 4 2023, doi: 10.3390/diagnostics13091624.

[16] S. Peng, W. Chen, J. Sun, and B. Liu, "Multi-Scale 3D U-Nets: An approach to automatic segmentation of brain tumor," International Journal of Imaging Systems and Technology, vol. 30, no. 1, pp. 5-17, 2020, doi: https://doi.org/10.1002/ima.22368.

[17] M. Ghaffari et al., "Automated post-operative brain tumour segmentation: A deep learning model based on transfer learning from pre-operative images," Magn Reson Imaging, vol. 86, pp. 28-36, Feb 2022, doi: 10.1016/j.mri.2021.10.012.

[18] A. Bianconi et al., "Deep learning-based algorithm for post-operative glioblastoma MRI segmentation: a promising new tool for tumor burden assessment," Brain Inform, vol. 10, no. 1, p. 26, Oct 6 2023, doi: 10.1186/s40708-023-00207-6.

[19] L. Guo et al., "A fuzzy feature fusion method for auto-segmentation of gliomas with multi-modality diffusion and perfusion magnetic resonance images in radiotherapy," Scientific Reports, vol. 8, no. 1, p. 3231, 2018/02/19 2018, doi: 10.1038/s41598-018-21678-2.

[20] Y. Zhu et al., "Semi-automatic segmentation software for quantitative clinical brain glioblastoma evaluation," Acad Radiol, vol. 19, no. 8, pp. 977-85, Aug 2012, doi: 10.1016/j.acra.2012.03.026.

[21] D. S. Chow et al., "Semiautomated volumetric measurement on postcontrast MR imaging for analysis of recurrent and residual disease in glioblastoma multiforme," AJNR Am J Neuroradiol, vol. 35, no. 3, pp. 498-503, Mar 2014, doi: 10.3174/ajnr.A3724.

[22] K. Zeng et al., "Segmentation of Gliomas in Pre-operative and Post-operative Multimodal Magnetic Resonance Imaging Volumes Based on a Hybrid Generative-Discriminative Framework," (in eng), Brainlesion, vol. 10154, pp. 184-194, Oct 2016, doi: 10.1007/978-3-319-55524-9_18.

[23] S. Hernandez et al., "Resection cavity auto-contouring for patients with pediatric medulloblastoma using only CT information," J Appl Clin Med Phys, vol. 24, no. 7, p. e13956, Jul 2023, doi: 10.1002/acm2.13956.

[24] S. Tian et al., "A Multicenter Study on Deep Learning for Glioblastoma Auto-Segmentation with Prior Knowledge in Multimodal Imaging," (in English), Int J Radiat Oncol, vol. 117, no. 2, pp. E488-E488, Oct 1 2023, doi: https://doi.org/10.1016/j.ijrobp.2023.06.2299.

[25] R. Meier et al., "Automatic estimation of extent of resection and residual tumor volume of patients with glioblastoma," (in eng), J Neurosurg, vol. 127, no. 4, pp. 798-806, Oct 2017, doi: 10.3171/2016.9.Jns16146.

[26] R. H. Helland et al., "Segmentation of glioblastomas in early post-operative multi-modal MRI with deep neural networks," Sci Rep, vol. 13, no. 1, p. 18897, Nov 2 2023, doi: 10.1038/s41598-023-45456-x.

[27] E. Ermis et al., "Fully automated brain resection cavity delineation for radiation target volume definition in glioblastoma patients using deep learning," Radiat Oncol, vol. 15, no. 1, p. 100, May 6 2020, doi: 10.1186/s13014-020-01553-z.

[28] K. K. Ramesh et al., "A Fully Automated Post-Surgical Brain Tumor Segmentation Model for Radiation Treatment Planning and Longitudinal Tracking," Cancers, vol. 15, no. 15, p. 3956, 2023. [Online]. Available: https://www.mdpi.com/2072-6694/15/15/3956.

[29] D. Jha, M. A. Riegler, D. Johansen, P. Halvorsen, and H. D. Johansen, "DoubleU-Net: A Deep Convolutional Neural Network for Medical Image Segmentation," presented at the 2020 IEEE 33rd International Symposium on Computer-Based Medical Systems (CBMS), 2020. [Online]. Available: https://doi.ieeecomputersociety.org/10.1109/CBMS49503.2020.00111.

[30] A. Celaya et al., "PocketNet: A Smaller Neural Network for Medical Image Analysis," IEEE Transactions on Medical Imaging, vol. 42, no. 4, pp. 1172-1184, 2023, doi: 10.1109/TMI.2022.3224873.

[31] P. Autee, S. Bagwe, V. Shah, and K. Srivastava, "StackNet-DenVIS: a multi-layer perceptron stacked ensembling approach for COVID-19 detection using X-ray images," Phys Eng Sci Med, vol. 43, no. 4, pp. 1399-1414, Dec 2020, doi: 10.1007/s13246-020-00952-6.

[32] P. Xiao et al., "A deep learning based framework for the classification of multi- class capsule gastroscope image in gastroenterologic diagnosis," Front Physiol, vol. 13, p. 1060591, 2022, doi: 10.3389/fphys.2022.1060591.

[33] H. Gunasekaran, K. Ramalakshmi, D. K. Swaminathan, A. J, and M. Mazzara, "GIT-Net: An Ensemble Deep Learning-Based GI Tract Classification of Endoscopic Images," Bioengineering (Basel), vol. 10, no. 7, Jul 5 2023, doi: 10.3390/bioengineering10070809.

[34] M. E. Moseley et al., "Diffusion-weighted MR imaging of anisotropic water diffusion in cat central nervous system," (in eng), Radiology, vol. 176, no. 2, pp. 439-45, Aug 1990, doi: 10.1148/radiology.176.2.2367658.

[35] F. Isensee et al., "Automated brain extraction of multisequence MRI using artificial neural networks," Human Brain Mapping, vol. 40, no. 17, pp. 4952-4964, 2019, doi: https://doi.org/10.1002/hbm.24750.

[36] B. C. Lowekamp, D. T. Chen, L. Ibanez, and D. Blezek, "The Design of SimpleITK," Front Neuroinform, vol. 7, p. 45, 2013, doi: 10.3389/fninf.2013.00045.

[37] L. R. Dice, "Measures of the Amount of Ecologic Association Between Species," Ecology, vol. 26, no. 3, pp. 297-302, 1945, doi: 10.2307/1932409.

[38] A. T. Kessler and A. A. Bhatt, "Brain tumour post-treatment imaging and treatment-related complications," (in eng), Insights Imaging, vol. 9, no. 6, pp. 1057-1075, Dec 2018, doi: 10.1007/s13244-018-0661-y.